\documentclass[conference]{IEEEtran}
\usepackage[T1]{fontenc}
\usepackage{lmodern}
\usepackage{hyperref}
\usepackage{url} 
\usepackage{cite} % Multiple citations in one~\cite command
\usepackage{pgfplotstable}
\usepackage{array} % Required for pgfplotstable dec sep align
\pgfplotsset{compat=1.12}
\usepackage{booktabs}
\usepackage{todonotes}
\usepackage{listings}

\usepackage{enumitem}
\setlist{parsep=0pt, listparindent=0.5cm}

\begin{document}
\title{A Comparison of Parallel Graph Processing Implementations}
\author{
	\IEEEauthorblockN{Samuel D. Pollard}
	\IEEEauthorblockA{Computer and Information Science\\
		University of Oregon \\
		Eugene, OR 97403\\
		Email: spollard@cs.uoregon.edu}
	\and
	\IEEEauthorblockN{Boyana Norris}
	\IEEEauthorblockA{Computer and Information Science\\
		University of Oregon \\
		Eugene, OR 97403\\
		Email: norris@cs.uoregon.edu}
}
\maketitle
\begin{abstract}
The rapidly growing number of large network analysis problems has led to the emergence of many parallel and distributed graph processing systems---one survey in 2014 identified over 80. Since then, the landscape has evolved; some packages have become inactive while more are being developed. Determining the best approach for a given problem is infeasible for most developers. To enable easy, rigorous, and repeatable comparison of the capabilities of such systems, we present an approach and associated software for analyzing the performance and scalability of parallel, open-source graph libraries. We demonstrate our approach on five graph processing packages: GraphMat, the Graph500, the Graph Algorithm Platform Benchmark Suite, GraphBIG, and PowerGraph using synthetic and real-world datasets. We examine previously overlooked aspects of parallel graph processing performance such as phases of execution and energy usage for three algorithms: breadth first search, single source shortest paths, and PageRank and compare our results to Graphalytics.
\end{abstract}

\section{Introduction}
Our research is motivated by the current state of parallel graph processing. Fields such as social network analysis \mbox{\cite{Kang:2011:BillionGraph}} and computational biology \mbox{\cite{Paylopoulos:2011:biograph}} require the analysis of ever-increasing graph sizes. The wide variety of problem domains is resulting in the proliferation of parallel graph processing frameworks. The most comprehensive survey, released in 2014, identified and categorized over 80 different parallel graph processing systems not even considering domain specific languages~\mbox{\cite{Doekemeijer:2015:GPFSurvey}}. 

An overarching issue among these systems is the lack of comprehensive comparisons. One possible reason is the considerable effort involved in getting each system to run: satisfying dependencies and ensuring data are correctly formatted can be time consuming tasks. Moreover, some systems are developed with large, multi-node clusters in mind while others only work with shared memory, single-node computers. An example of the former is GraphX~\mbox{\cite{Xin:2013:GraphX}}, which incurs some overhead while gaining fault tolerance and an example of the latter is Ligra~\mbox{\cite{Shun:2013:Ligra}}, a framework requiring a shared-memory architecture. Additionally, there is a growing number of systems designed to run on GPUs \mbox{\cite{Zhong:2014:Medusa}}.
% Ligra's motivation for shared memory is the claim that `` the largest publicly available real-world graphs all fit in shared memory.''

From a graph algorithm user's point of view, optimizations for each system may not be apparent. For example, the Parallel Boost Graph Library (PBGL)~\mbox{\cite{Gregor:2005:PBGL}} provides generic implementations of their algorithms and the programmer must provide the template specializations. Optimal data structures may differ across graphs and must be determined by the programmer.

We aim to simplify the decision making process by providing an easy-to-use framework for consistently and fairly evaluating the performance of different algorithms. To illustrate the use of this framework, we include results from our experiments on both real-world and synthetic datasets. We take inspiration from the Graph500 benchmark~\mbox{\cite{Murphy:2010:Graph500}}, which clearly specifies every step of the breadth first search (BFS) algorithm and how it should be timed. The Graph500 can be used to rank systems fairly on a single, well-defined benchmark. In this context, \emph{system} is defined as the hardware, operating system, middleware, and algorithmic implementation which accomplishes a certain task, in this case BFS.

While leaderboards can indicate computational milestones, they are not particularly useful for the average user. Reference implementations, however, are critical for advancement; the complexity of today's hardware and operating systems requires empirical evidence to assess an algorithm's performance. In addition, there is motivation to define basic building blocks for graph computations~\mbox{\cite{GABB16, Buluc:CombBLAS:2011}}. However, new algorithm designers and users alike have no easy way to accurately determine what constitutes a high-performance implementation.

To address these issues, we introduce \emph{easy-parallel-\mbox{graph-\textasteriskcentered}},\footnote{Available at \url{https://github.com/HPCL/easy-parallel-graph}} a framework which simplifies the installation, comparison, and performance analysis of the three most widely implemented graph algorithm building blocks: breadth first search (BFS), single source shortest paths (SSSP), and PageRank (PR). With this framework we select a small number of graph processing libraries over which we homogenize all aspects of execution to compare performance fairly. Another contribution of this paper is a demonstration of the use of this framework to analyze the performance, energy consumption, and scalability of specific algorithm implementations. This enables algorithm designers to select high performance reference implementations with which they can compare their improvements. To be clear, we are not proposing a new benchmark suite or providing any new reference implementations. Instead, we introduce a method for detailed and fair comparison of existing graph software packages without requiring the user to be familiar with them.

Section~\ref{sec:relatedwork} describes previous performance analysis of parallel graph algorithms. Section~\ref{sec:meth} describes the architecture of \emph{easy-parallel-graph-\textasteriskcentered} and the algorithms, datasets, and graph processing implementations it uses. Section~\ref{sec:perf} presents performance analysis and discusses possible explanations for the results. Section~\ref{sec:fw} describes future directions for this research and is followed by the conclusion.

\section{Related Work}\label{sec:relatedwork}
Most performance analysis of graph processing systems come from the empirical results of each new library designer's publications. For example, GraphMat~\mbox{\cite{Sundaram:2015:GraphMat}}, PowerGraph~\mbox{\cite{Gonzalez:2012:Powergraph}}, and GraphBIG~\mbox{\cite{Nai:2015:Graphbig}} present a comparison of performance between their system and a selection of other software packages and in general any new approach will compare itself to existing implementations. We assume the performance results which accompany a new library are at risk of bias because the authors typically know their own package better than the competition and as such are aware of more optimizations. For a collection of references to the original papers of each new library or framework, see \mbox{\cite{Doekemeijer:2015:GPFSurvey}}. Instead, we focus on related research which provide analysis of existing software.

Nai et al.~\mbox{\cite{Nai:2016:architectural}} provide a detailed performance analysis using GraphBIG as their reference implementation. Their approach also compares GraphBIG to the GraphLab and Pregel execution models (Gather Apply Scatter). Their analysis considers architectural performance measurements such as cache miss rates to measure bottlenecks and computational efficiency for a variety of datasets. While incredibly thorough, these analyses focus on covering a wide range of datasets and appear difficult to apply to a new approach. For example, GraphMat reduces computation to sparse matrix operations which may not suffer from the same memory bottlenecks indicated in~\mbox{\cite{Nai:2016:architectural}}.

Research by Song et al.~\mbox{\cite{Song:2016:LoadBalancing}} and LeBeane et al.~\mbox{\cite{LeBeane:2015:GraphHetero}} considers the important problem of partitioning on heterogeneous architectures and provides profiling to reduce runtime, improve load balance, and reduce energy consumption.  Our framework instead focuses on shared memory packages. 

Beyond the aforementioned performance results, a number of reports focus on performance analysis not tied to a particular software package. Satish et al.~\mbox{\mbox{\cite{Satish:2014:NavigatingGraph}}} analyze the performance of several systems on datasets on the order of $30$ billion edges and~\mbox{\cite{Lu:2014:ExperimentalEval}} uses six real-world datasets and focuses on the vertex-centric programming model. These implementations are hand tuned and provide recommendations for future improvements.

\begin{table}
	\caption{Graphalytics: tabulated sample run times (seconds) with 32 threads. Just one run per experiment is performed. Below the table is an excerpt from the GraphMat log file.}
	\label{tab:graphalytics-realworld}
	\vspace{-10pt}
	\begin{center}
		% cit-Patents: (3 774 768 vertices, 16 518 948 edges)
		% dota-league (61 170 vertices, 50 870 313 edges)
		\begin{tabular}{l|c|c|c|c|c|c}
			%% BN: Commented out because it's redundant with the caption
			%\multicolumn{5}{l}{\large{\emph{Graphalytics}}}  \\
			GraphBIG & BFS  & CDLP  &   LCC   &  PR &  SSSP & WCC \\ \hline
			cit-Patents & 0.8 & 11.8 & 15.5  &  4.5 & N/A  & 1.3 \\
			dota-league & 1.1 & 3.9 & 1073.7 &  {\bf 2.6} & 3.0 & 1.0 \\
			\multicolumn{7}{l}{} \\
			PowerGraph & BFS  & CDLP  &   LCC   &  PR &  SSSP & WCC \\ \hline
			cit-Patents & 13.8 & 30.1 & 23.9 & 18.8 & N/A & 22.1 \\
			dota-league & 25.6 & 31.2 & 458.1 & 26.7 & 28.9 & 22.9 \\
			% graph500-22 43.0 s 55.6 s 299.8 s 46.4 s N/A    40.5 s
			\multicolumn{7}{l}{} \\
			GraphMat & BFS  & CDLP  &   LCC   &  PR &  SSSP & WCC \\ \hline
			cit-Patents &  7.5 &  20.1 & 9.8 &  8.1 & N/A  & 6.6 \\
			dota-league &  2.7 &  21.2 & 239.7 & {\bf 6.3} & 9.4 & 6.9 \\
		\end{tabular}
	\end{center}

	\noindent Timing results (for GraphMat PageRank on dota-league)
	\begin{itemize}
		\item Finished file read of dota-league. time: 2.65211
		\item load graph: 5.91229 sec
		\item initialize engine: 8.32081e-05 sec
		\item run algorithm 1 (count degree): 0.0555639 sec
		\item run algorithm 2 (compute PageRank): 0.149445 sec
		\item print output: 0.0641179 sec
		\item deinitialize engine: 0.00022006 sec
	\end{itemize}
\end{table}

The most prominent example of a graph processing performance analysis tool not tied to a particular implementation is Graphalytics~\mbox{\cite{Capota:2015:Graphalytics}}. Similar to our work, Graphalytics also automates the setup and execution of graph packages for performance analysis. Graphalyics relies on Apache Maven and Java to wrap the execution of each graph processing software package.
%---Graphalytics calls these platforms---and requires configuration files for each dataset.\todo[inline]{What am I saying with this?}
%Moreover, the complexity of Graphalytics can obfuscate the true behavior of the program.

Graphalytics generates an HTML report listing the runtimes for each dataset and each algorithm. Table~\ref{tab:graphalytics-realworld} shows the results from a typical experiment run by Graphalytics.

However, perusal of the log files reveals a different story. The bullets shown below Table~\ref{tab:graphalytics-realworld} summarize the output from GraphMat itself. Graphalytics reports a 6.3 second runtime but 2.7 seconds of that time GraphMat is simply reading the input file from disk. However, the GraphBIG timing (2.6 seconds) does not include the time to read the dota-league file. \textbf{If the time to read in the text file was ignored then GraphMat would complete nearly twice as quickly.} To call this a fair comparison is dubious at best when certain phases of execution are omitted for some packages but not others.

%We use GraphMat, the single source shortest paths algorithm, the dota-league dataset from the Game Trace Archive\mbox{\cite{Guo:2012:GTA}}, and the default settings as an example here.
%source vertex: 55713
%num. threads: 64
%Starting file read of /home/users/spollard/graphalytics/graphalytics-platforms-graphmat/intermediate/dota-league.e_weight..mtx
%Got graph with m=61170  n=61170 nnz=101740626
%Finished file read of /home/users/spollard/graphalytics/graphalytics-platforms-graphmat/intermediate/dota-league.e_weight..mtx, time: 0.94948
%Finished setting ids, time: 0.077502
%Starting sort
%Finished sort, time: 0.767248
%Finished setting edge pointers, time: 0.00017
%Starting build_dcsc
%Finished build_dcsc, time: 0.162558
%Finished setting ids, time: 0.07698
%Starting sort
%Finished sort, time: 1.20438
%Finished setting edge pointers, time: 0.000169
%Starting build_dcsc
%Finished build_dcsc, time: 0.156858
%Completed reading A from memory in 2.522761 seconds.
%Completed reading A from file in 3.481659 seconds.
%Completed 19 iterations
%Timing results:
%- load graph: 3.48167 sec
%- initialize engine: 3.60012e-05 sec
%- run algorithm: 0.168526 sec
%- print output: 0.105846 sec
%- deinitialize engine: 6.00815e-05 sec
%03:17:45.114 [INFO ] Benchmarked algorithm "Single source shortest paths" on graph "dota-league".
%03:17:45.114 [INFO ] Benchmarking algorithm "Single source shortest paths" on graph "dota-league" succeed.
%03:17:45.114 [INFO ] Benchmarking algorithm "Single source shortest paths" on graph "dota-league" took 3830 ms.
With a plugin to Graphalytics called Granula~\mbox{\cite{Ngai:2015:Granula}}, one can explicitly specify a performance model to analyze specific execution behavior such as the amount of communication or runtime of particular kernels of execution. This requires in-depth knowledge of the source code and execution model in addition to expertise with the Granula API but allows detailed performance analysis and automatic execution and compilation of performance results.\footnote{An example of Granula can be seen at \url{http://bit.ly/granula-example}}.
%\url{https://github.com/tudelft-atlarge/graphalytics-platforms-graphx/tree/master/granula-model-graphx}}.

Beyond this plethora of performance data, the Graph Algorithm Platform Benchmark Suite (GAP)~\mbox{\cite{Beamer:2015:GAPBench}}, GraphBIG~\mbox{\cite{Nai:2015:Graphbig}}, and the Graph500~\mbox{\cite{Murphy:2010:Graph500}} consider themselves reference implementations or benchmark suites with which other implementations can be compared. If no fewer than three software packages call themselves, ``reference implementations,'' which one are we to trust? Our belief is that choosing a single reference implementation cannot capture the complexities inherent in graph processing.

% The systems described in~\mbox{\cite{Doekemeijer:2015:GPFSurvey}} operate with a wide range of parallelism paradigms and target architectures such as GPU~\mbox{\cite{Zhong:2014:Medusa, Kang:2009:Pegasus}}, shared memory CPU~\mbox{\cite{Shun:2013:Ligra, kyrola:2012:Graphchi, Nguyen:2013:Galois}}, a combination of CPU and GPU~\mbox{\cite{Gharaibeh:2012:Totem}}, distributed filesystem based approaches~\mbox{\cite{Xin:2013:GraphX}}, and distributed memory with MPI~\mbox{\cite{Gregor:2005:PBGL}}.

% Beyond the systems described by Doekemeijer and Varbanescu, the problem has compounded with the addition of even more proprietary and open source projects such as~\mbox{\cite{Cheramangalath:2015:Falcon, Perez:2015:Ringo}}, distributed memory approaches such as~\mbox{\cite{Hong:2015:PGX}}. domain-specific languages~\mbox{\cite{Hong:2012:GreenMarl}}, distributed database querying,~\mbox{\cite{Rodriguez:2015:Gremlin}}, as well as novel communication schemes~\mbox{\cite{Edmonds:2013:ActiveMessages}}. At the outset, this plethora of choices makes the question, ``which system is the best for my problem?'' daunting.

% In addition to libraries with associated APIs there has also been a propagation of ``reference implementations'' which implement the most common graph algorithms such as~\mbox{\cite{Beamer:2015:GAPBench, Nai:2015:Graphbig}}. Thus, even selecting a standard and a benchmark over which to compare various implementations is nontrivial. To quote Andrew Tanenbaum, ``The nice thing about standards is that you have so many to choose from.''

\section{Methodology}\label{sec:meth}

\begin{figure*}
	\centering
	% trim=left lower right upper
	\includegraphics[width=\linewidth, trim=0 144pt 0pt 156pt, clip]{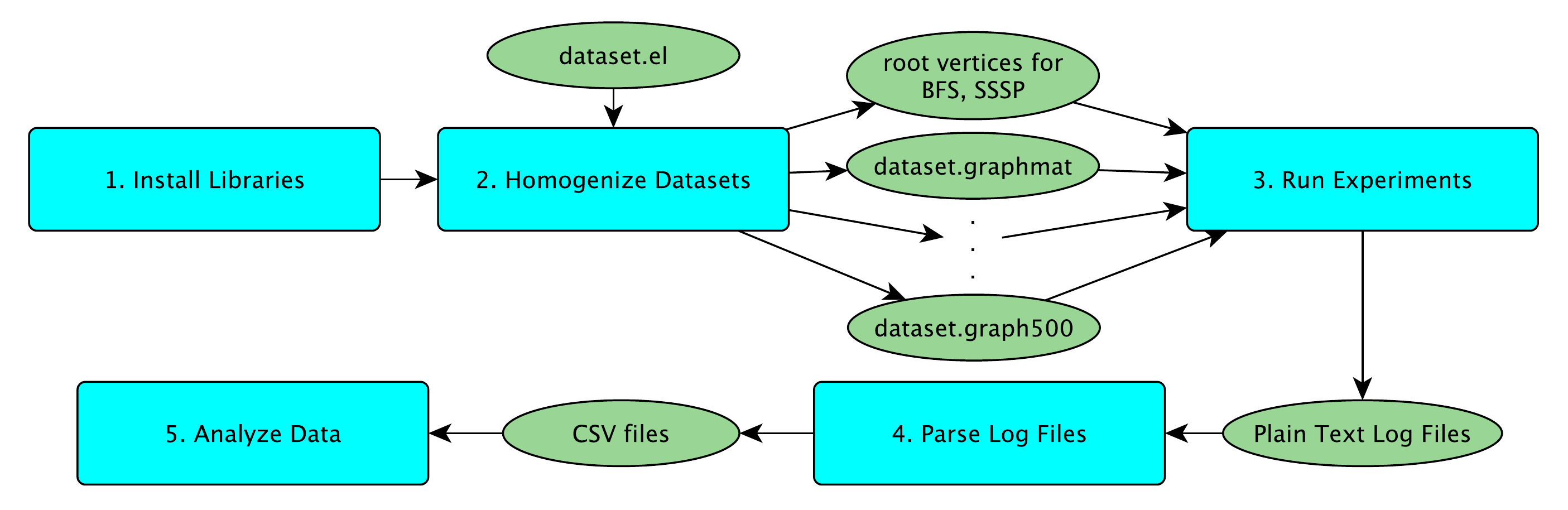}
	\caption{Overview of \emph{easy-parallel-graph-\textasteriskcentered}. Each cyan box corresponds to a single shell script in the package while the green ellipses correspond to generated files.}
	\label{fig:epg-overview}
\end{figure*}

Our contribution strikes a balance between the highly detailed analysis such as those presented by Nai~\mbox{\cite{Nai:2016:architectural}} (greater depth of analysis) and Graphalytics which compares only the runtimes of a greater number of software packages and algorithms (greater breadth of analysis). This middle-ground simplifies our interface which in turn allows users to perform their own analysis on their own datasets with greater ease than existing frameworks.

In addition, we analyze  power usage, energy consumption, and details such as the number of iterations for PageRank or the time to construct the graph data structures. Moreover, our approach requires little knowledge of the inner workings of each system; the data are collected by either parsing log files (for execution time) or hardware counter sampling of model-specific registers (for power). This allows arbitrary datasets to be included as long as they are in the same format as those in the Stanford Network Analysis Project (SNAP) datasets~\mbox{\cite{snapnets}}.

Our framework breaks the process of characterizing performance into five principal phases shown in \figurename~\ref{fig:epg-overview}, each of which requires no more than a single shell command. The five phases are as follows.
\begin{enumerate}
	\item Installing modified, stable forks of each software package to ensure homogeneity.
	\item Given a synthetic graph size or a real-world graph file, generate the files necessary to run each software package.
	\item Given a graph and the number of threads, run each algorithm using each software package multiple times.
	\item Parse through the log files to compress the output into a CSV.
	\item Analyze the data using the provided R scripts to generate plots.
\end{enumerate}

\subsection{Installing Libraries}
The libraries are stable forks of the given repositories which are configured to ensure the experiments execute in the same manner.

%% How do we address these issues?
Our experiments use the author-provided implementations with modifications only to insert performance analysis hooks or to ensure homogeneous stopping criteria; we assume the developers of each system will provide the best performing implementation. While this limits the scope of the experiments it mitigates the bias inherent in our programming skills in addition to the bias of library designers' performance analysis; a given library designer will understand his or her source code better than any other implementation.

\subsection{Datasets}
Homogenizing the datasets creates copies of the graph files and auxiliary files in various formats. This is both to ensure they are correctly formatted for each system and to speed up file I/O whenever possible by using the library designer's serialized data structure file formats.

We refer to a graph's \emph{scale} when describing the size of the graph. Specifically, a graph with scale $S$ has $2^S$ vertices. For example, many of our experiments were performed on graphs with $2^{22} = 4,194,304$ vertices and an average of 16 edges per vertex. We measure parallel efficiency and speedup by varying the number of threads from one to the total number of threads available on our research server, 72.

Each experiment uses 32 roots per graph. As with the Graph500, each root is selected to have a degree greater than 1. For PageRank, we simply run the algorithm 32 times. We plot many of the results as box plots with an implied 32 data points per box.

When possible, we measure the dataset construction time as the time to translate from the unstructured file data in RAM to the graph representation on which the algorithm can be performed. This is not possible for PowerGraph and GraphBIG because they read in the input file and build a graph simultaneously.

We use the Graph500 synthetic graph generator which creates a Kronecker graph~\mbox{\cite{Leskovec:2010:Kronecker}} with initial parameters of $A = 0.57, B = 0.19, C = 0.19,$ and $D = 1-(A+B+C) = 0.05$ and set the average degree of a vertex as 16. Hence, a Kronecker graph with scale $S$ has $2^S$ vertices and approximately $16 \times 2^S$ edges. Kronecker graphs are a generalization of RMAT graphs.

Part of the Graphalytics results in Table~\ref{tab:graphalytics-realworld} were performed on the Dota-League dataset. We use this dataset for other experiments and as well. It contains 61,670 vertices and 50,870,313 edges and models interactions between players in the online video game Defense of the Ancients. This was sourced from the Game Trace Archive\mbox{\cite{Guo:2012:GTA}} and is modified for Graphalytics\footnote{This dataset is available at \url{https://atlarge.ewi.tudelft.nl/graphalytics/}.}. This dataset is useful because it is both weighted and more dense than the usual real-world dataset with an average out-degree of 824.

We also present results for the \verb|cit-Patents| dataset from SNAP \mbox{\cite{snap-cit-patents}}. This is a widely-used network of citations from the National Bureau of Economic Research (NBER) and is less dense than Dota-League with 3,774,768 vertices and 16,518,948 edges. We stress that though these two datasets are presented, any network in the SNAP data format\footnote{A file in the SNAP format consists of one edge per line, with vertices separated by whitespace and lines which begin with \# are comments.} can be used in \mbox{\emph{easy-parallel-graph-\textasteriskcentered}}.

\subsection{Graph Processing Systems}

This study explores four shared memory parallel graph processing frameworks and one distributed memory framework operating on a single node (Powergraph). The first three are so-called ``reference implementations'' while the remaining two are included because of their performance and popularity. Other popular libraries such as the Parallel Boost Graph Library~\mbox{\cite{Gregor:2005:PBGL}} are not considered here because the authors do not provide reference implementations. The frameworks are as follows:
\begin{enumerate}
	\item The Graph500~\mbox{\cite{Murphy:2010:Graph500}}, the canonical BFS benchmark which consists of a specification and reference implementation. We use a modified version most similar to 2.1.4. The Graph500 provides reference implementations for MPI and OpenMP but since our framework focuses on shared memory we use only the OpenMP version. The Graph500 uses a compressed sparse row (CSR) representation.
	\item The Graph Algorithm Platform (GAP) Benchmark Suite~\mbox{\cite{Beamer:2015:GAPBench}}, a set of reference implementations for shared memory graph processing. The author of GAP has contributed to the Graph500 so the BFS implementations are similar. Additionally, GAP uses OpenMP to achieve parallelism and uses a CSR representation.
	\item GraphBIG~\mbox{\cite{Nai:2015:Graphbig}} benchmark suite. We consider only the shared memory solutions but GraphBIG also provides GPU benchmarks. GraphBIG uses a CSR representation for graphs and OpenMP for parallelism.
	\item GraphMat~\mbox{\cite{Sundaram:2015:GraphMat}}, a library and programming model along with reference implementations of common algorithms. GraphMat uses a doubly-compressed sparse row representation and OpenMP for parallelism.
	\item PowerGraph~\mbox{\cite{Gonzalez:2012:Powergraph}}, a library and programming model for distributed (and shared memory) graph-parallel computation. PowerGraph includes additional toolkits for tasks such as clustering and computer vision. Parallelism is achieved via a combination of OpenMP and light-weight, user-level threads called fibers. Powergraph uses a novel storage scheme on top of CSR.
\end{enumerate}

Our approach is not specific or limited to these graph packages and can be extended to others. When comparing to Graphalytics, we use the most recent stable release, v0.3.

\subsection{Algorithms}\label{sec:algs}

We consider three parallel algorithms: Breadth First Search (BFS), Single Source Shortest Paths (SSSP), and PageRank, though not all algorithms are implemented on all systems. We inspected the source code of the surveyed parallel graph processing systems to ensure the same phases of execution are measured across differing execution and programming paradigms. 

We select BFS because the canonical performance leaderboard for parallel graph processing is the Graph500~\mbox{\cite{Murphy:2010:Graph500}}. One strength of the Graph500 is it provides standardized measurement specifications and dataset generation. The primary drawback with the Graph500 is it measures a single algorithm.

Our work aims to add similar rigor to other graph algorithms by borrowing heavily from the Graph500 specification. The Graph500 Benchmark 1 (``Search'') is concerned with two kernels: the creation of a graph data structure from an unsorted edge list stored in RAM and the actual BFS\footnote{For a complete specification, see \url{http://graph500.org/specifications}}. We run the BFS using 32 random roots with the exception of PowerGraph which doesn't provide an reference implementation of BFS in its toolkits.

We select SSSP because of the straightforward extension from BFS; we need not modify the graph and can use the same the root vertices from BFS. Furthermore, SSSP is used as a building block for other graph algorithms such as Betweenness Centrality.

PageRank is selected because of its popularity; most libraries provide reference implementations. One challenge with using PageRank is the stopping criterion; all implementations have been modified to use $|| p_t - p_{t-1}||_1$ (the absolute sum of differences) where $p_t$ is the page rank at step $t$. Verification of the PageRank results is beyond the scope of this paper, although this may explain some of the large performance discrepancies.

This approach is not specific to a particular algorithm; measuring the execution time, data structure construction time, and power consumption can be applied easily to other algorithms. We select some representative algorithms here to demonstrate the process, which we hope will motivate others to use
the framework to add more algorithms.

\subsection{Parsing and Data Analysis}
Our example analyzes data using the R programming language and shell (Bash and AWK) parsers. The example workflow provided in our framework yields the figures and tables generated in this paper.

\subsection{Machine Specifications}
We performed experiments on our 36-core (72 thread) Intel Haswell server with 256GB DDR4 RAM, with two Intel Xeon E5-2699 v3 CPUs. The operating system is GNU/Linux version 4.4.0-22. Our code was compiled with GCC version 4.8.5 with the exception of GraphMat which was compiled with ICPC version 17.0.0.
%\begin{center}
%	% For arya I deleted Max RAM Freq	2133MHz
%	\pgfplotstabletypeset[
%	header=false,
%	col sep=tab,
%	string type,
%	every head row/.style={output empty row, before row=\bottomrule},
%	columns/0/.style={column type={|p{1in}|}},
%	columns/1/.style={column type={p{3.2in}|}},
%	every last row/.style={after row=\toprule},
%	]{../report/specs.csv}
%	%\label{tab:specs}
%\end{center}

\section{Performance Analysis}\label{sec:perf}

We analyze the performance of the algorithms described in Sec.~\ref{sec:algs} in terms of execution time, scalability over multiple threads, and power and energy consumption. All figures except those measuring scalability (Figs.~\ref{fig:bfs-speedup} and~\ref{fig:bfs-efficiency}) use 32 threads.

\subsection{Runtime for Synthetic Graphs}
\begin{table}
	\caption{Graphalytics on the same Kronecker graph with scale $22$ as used in other experiments. Performance results are in seconds with 32 threads. Community detection uses label propagation.}
	\centering
	\begin{filecontents*}{runtime.csv}
,graphmat,openg,powergraph
Community Detection,45.8,7.4,55.6
PageRank,8.9,4.7,46.4
Local Clustering Coeff.,401.0,1802.7,299.8
Weakly Conn. Comp.,7.4,2.4,40.5
BFS,10.3,1.8,43.0
\end{filecontents*}
\pgfplotstabletypeset[
	col sep=comma,
	columns={[index]0,graphmat,openg,powergraph},
	every head row/.style={after row=\midrule},
	columns/0/.style={string type, column type={l|}, column name={\large{\emph{Graphalytics}}}},
	columns/graphmat/.style={
		column name={GraphMat},
		string replace={0}{},
		dec sep align,
		empty cells with={N/A}
	},
	columns/openg/.style={
		column name={GraphBIG},
		dec sep align},
	columns/powergraph/.style={
		column name={PowerGraph},
		dec sep align}
	]{runtime.csv}
	\label{tab:graphalytics}
\end{table}

\begin{figure}
	\centering
	\begin{minipage}{0.48\linewidth}
		\includegraphics[width=\linewidth, trim=0pt 18pt 18pt 0pt, clip]{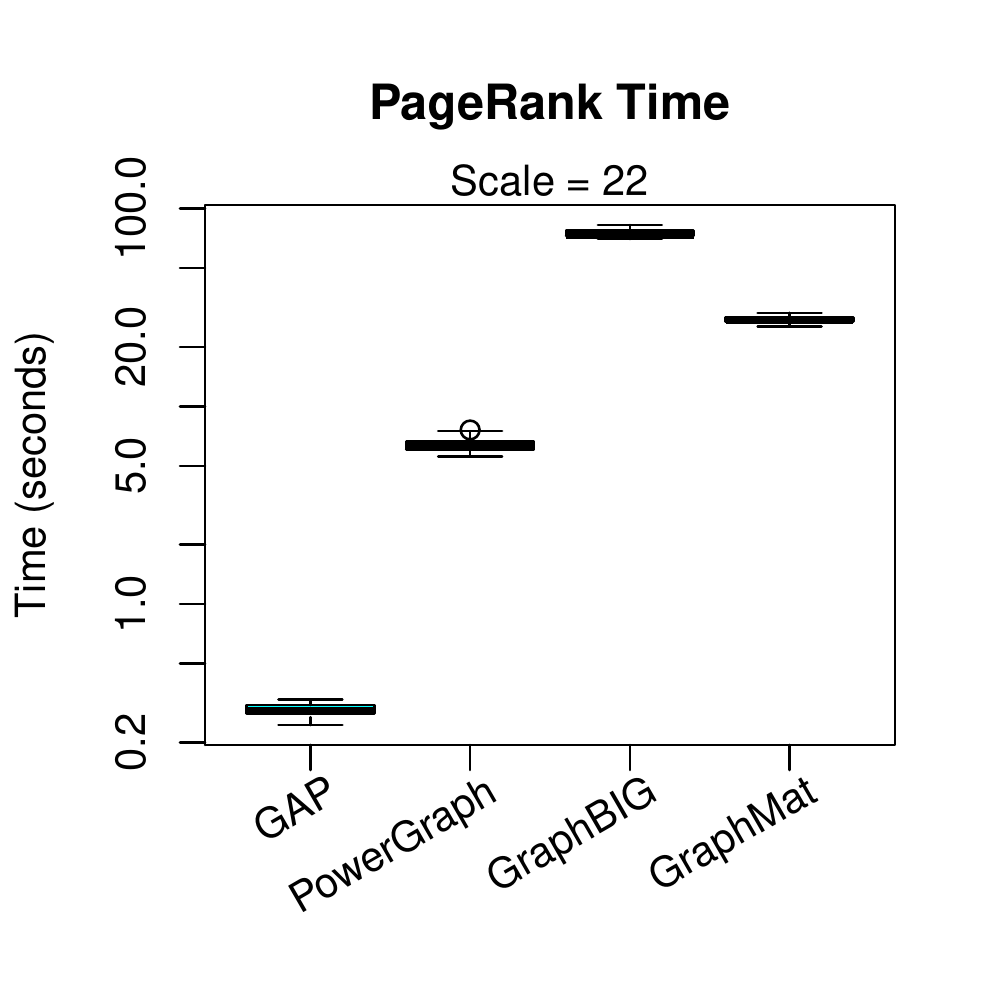}
	\end{minipage}
	\begin{minipage}{0.48\linewidth}
		\includegraphics[width=\linewidth, trim=0 18pt 18pt 0pt, clip]{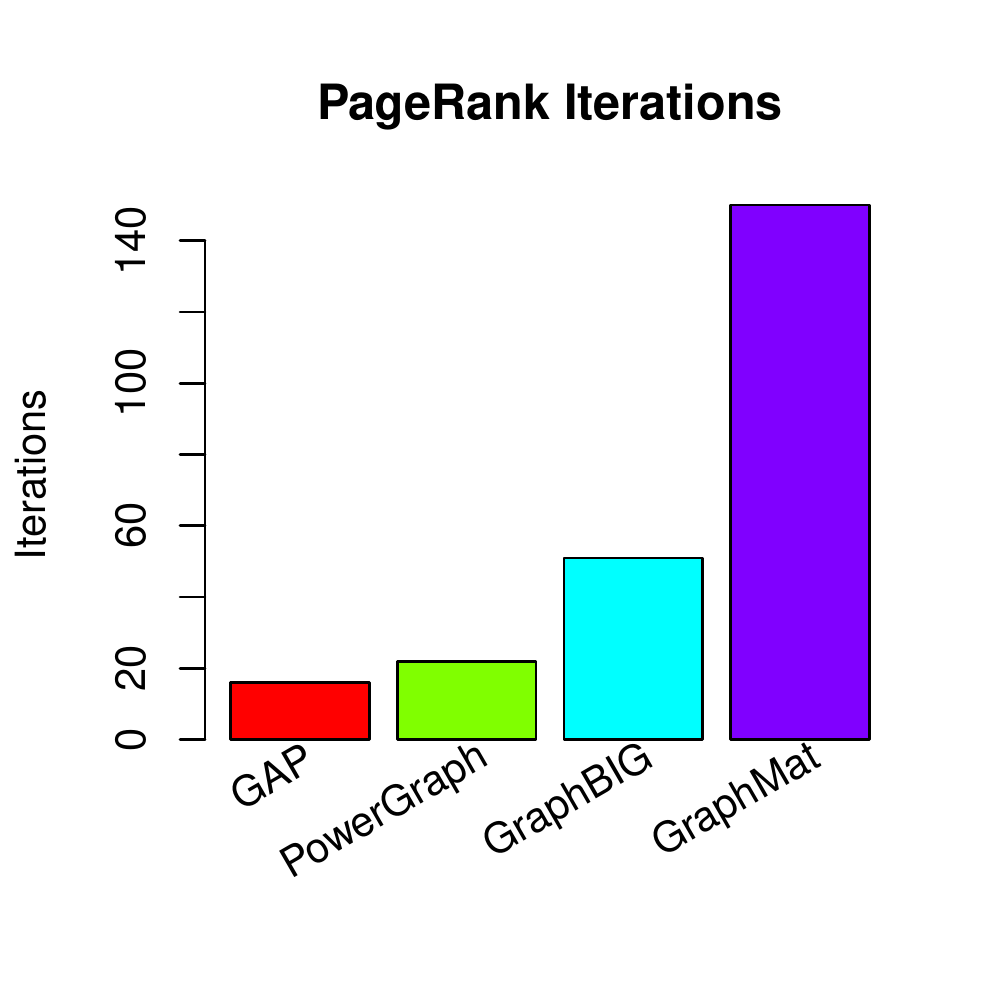}
	\end{minipage}
	\caption{The $y$-axis is logarithmic only for the left figure. GraphMat continues to run until none of the vertices' ranks change. For the others, we use the stopping condition that the sum of the changes in the weights is no more than $6 \times 10^{-8}$, or approximately machine epsilon for single precision floating point numbers.}
	\label{fig:pr}
\end{figure}

In Table~\ref{tab:graphalytics} we show the results from running Graphalytics on a Kronecker graph of scale $22$ (4,194,304 vertices and approximately $16 \times 2^{22} \approx 33,500,000$ edges). An explanation of each algorithm is given in~\mbox{\cite{Iosup:2016:Graphalyticstech}}. Graphalytics by default does not perform SSSP on unweighted, undirected graphs. The discrepancy between PageRank values in Table~\ref{tab:graphalytics} and Fig.~\ref{fig:pr} is a result of the differing stopping criterion and the aforementioned inconsistency of Graphalytics's performance collection scheme.

\begin{figure}
	\centering
	\begin{minipage}{0.48\linewidth}
		\includegraphics[width=\linewidth, trim=0 36pt 18pt 0, clip]{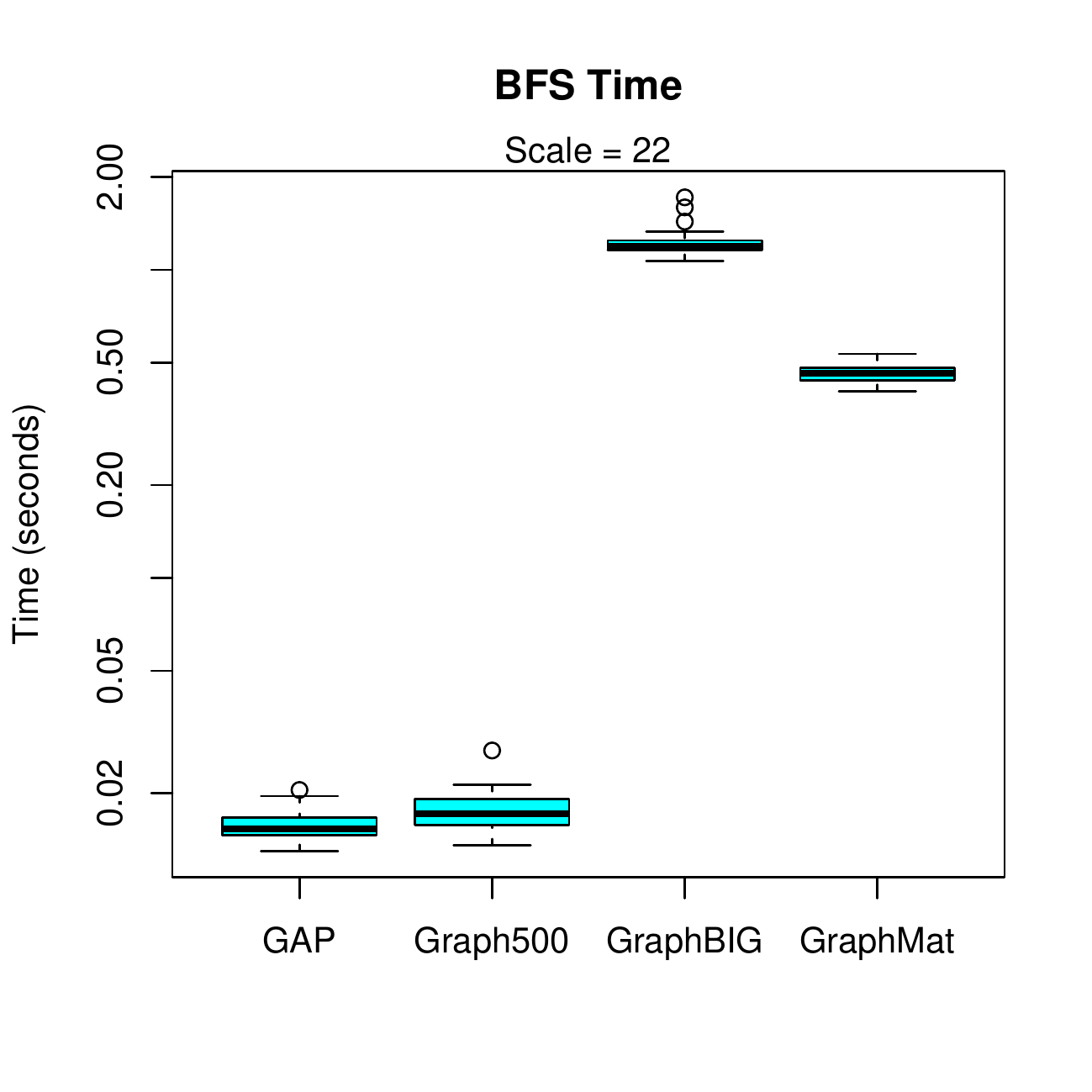}
	\end{minipage}
	\begin{minipage}{0.48\linewidth}
		\includegraphics[width=\linewidth, trim=0 36pt 18pt 0, clip]{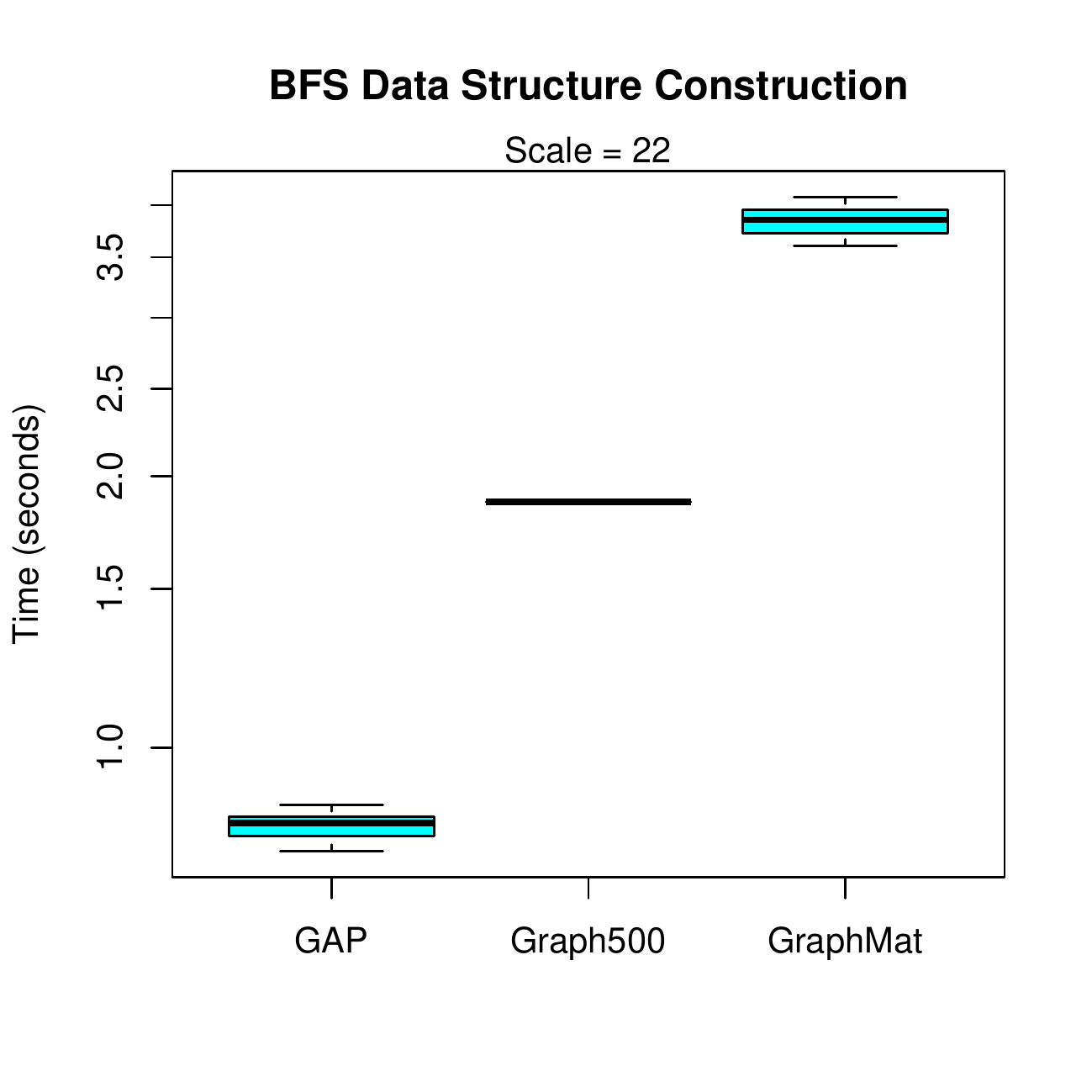}
	\end{minipage}
	\caption{The $y$-axes are logarithmic. The left box plot shows the time to compute BFS on 32 random roots while the right plot shows the times to construct the graph for each system. The Graph500 only constructs its graph once. GraphBIG reads in the file and generates the data structure simultaneously so is omitted.}
	\label{fig:bfs-time}
\end{figure}

\begin{figure}
	\centering
	\begin{minipage}{0.59\linewidth}
		\includegraphics[width=\linewidth, trim=0 36pt 18pt 0, clip]{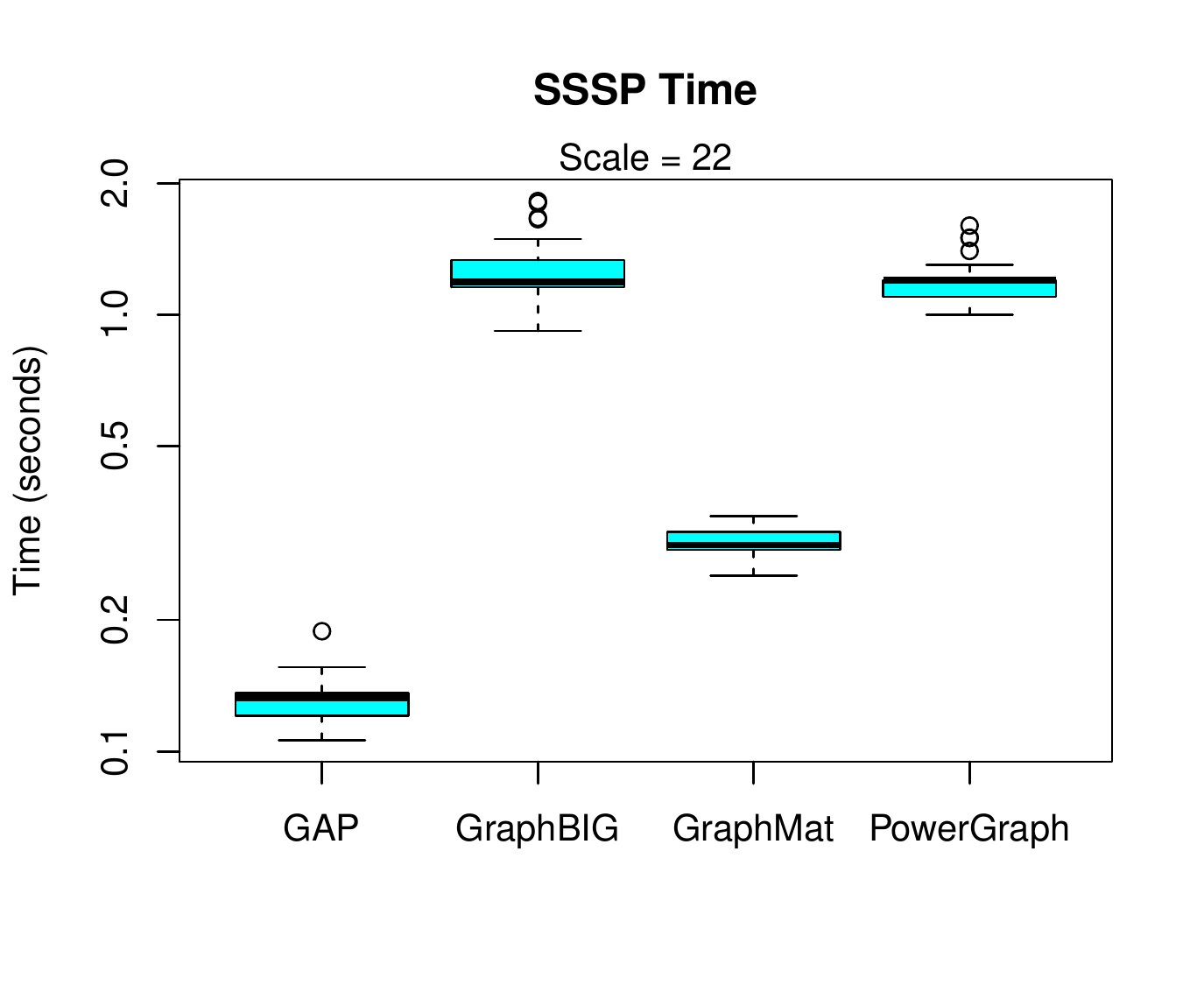}
	\end{minipage}
	\begin{minipage}{0.365\linewidth}
		\includegraphics[width=\linewidth, trim=0 36pt 18pt 0, clip]{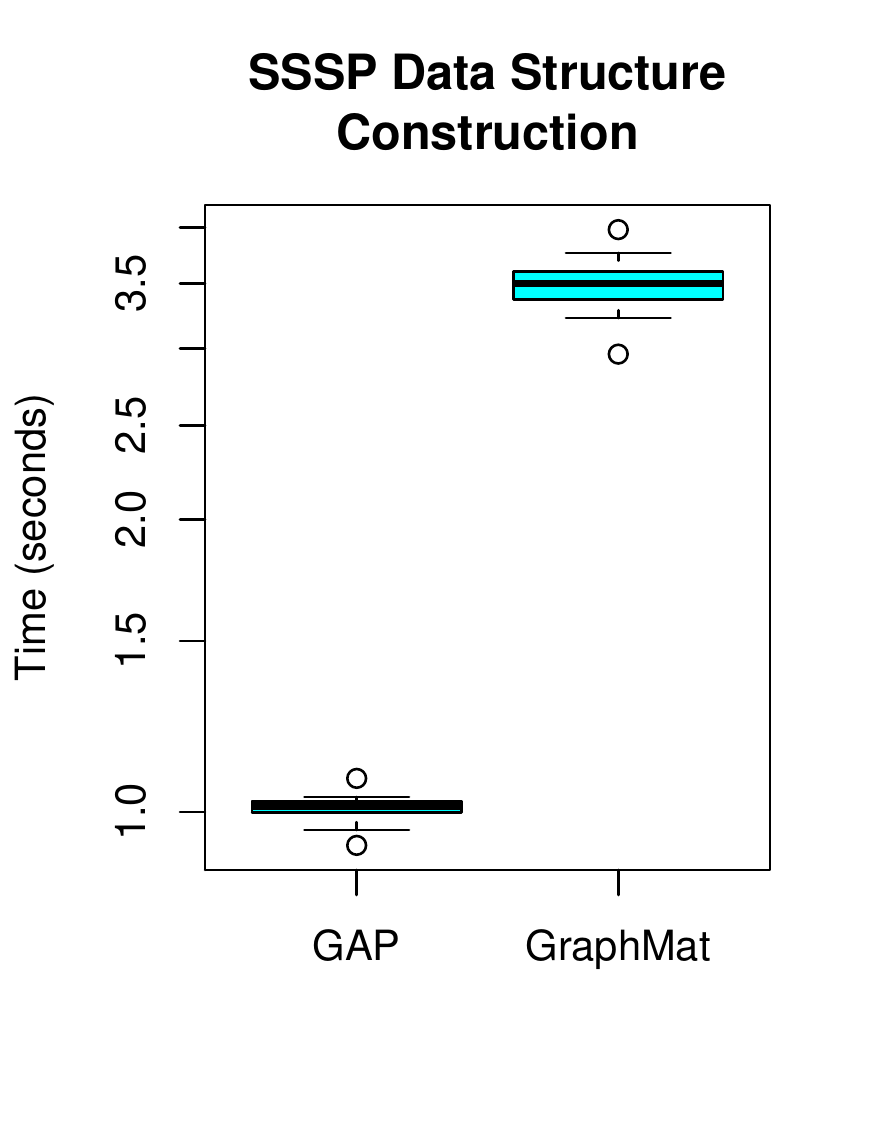}
	\end{minipage}
	\caption{The $y$-axes is logarithmic. The left box plot shows the time to compute the SSSP starting at the same 32 roots as Fig.~\ref{fig:bfs-time}. Both PowerGraph and GraphBIG construct their data structures at the same time as they read the file.}
	\label{fig:sssp-time}
\end{figure}

In contrast, our results for the Kronecker graph of scale $22$ are presented in Figs.~\ref{fig:bfs-time},~\ref{fig:sssp-time}, and~\ref{fig:pr}.

% \todo{What are the likely culprits for performance differences---measure cache miss rates. Try to find something which correlates with performance. Resource stalls will work easily (not as interesting) Cycles per instruction? Automate with perfexplorer or autoperf}
% Graphalytics also outputs MTEPS or millions of traversed edges per second. However, the graphalytics version does not make sense in all cases: for example, computing the local clustering coefficient involves traversing each edge multiple times (proportional to the sparsity of the graph), while BFS traverses each edge exactly once, and the number of edges traversed with PageRank depends on the connectivity of the graph and the number of iterations.

Figures~\ref{fig:bfs-time} and \ref{fig:sssp-time} show performance results for a Kronecker graph with scale $22$. The box plots give an idea of the runtime distributions. There is less variance in the runtimes of SSSP (between 0.1 and 1.7 seconds) compared to BFS (0.01 and 1.7 seconds) but GAP is the clear winner in both cases. The data structure construction times for GAP and GraphMat are consistent; in both cases the platforms create the same data structure for both algorithms. These results are consistent with~\mbox{\cite{Sundaram:2015:GraphMat}} which lists GraphMat as more higher performing than PowerGraph in SSSP.

The behavior of PageRank is slightly different. As with SSSP and BFS, the GAP Benchmark Suite is the fastest but it also requires the fewest iterations. We attempt to define similar stopping criteria for each system, but GraphMat executes until no vertices change rank; effectively its stopping criterion requires the $\infty$-norm be less than machine epsilon. This could account for the increased number of iterations. We adjusted the other systems to use $\sum_{k=1}^{n} |p_k^{(i)} - p_k^{(i-1)}| < \epsilon $ as the stopping criterion, where $i$ is the iteration and $n$ is the number of vertices. We use $\epsilon = 6 \times 10^{-8}$ because this value is approximately machine epsilon for a single precision floating-point number to make the stopping criteria for all implementations as similar as possible. However, with GraphMat there is no computation of $|p_k^{(i)} - p_k^{(i-1)}|$.

Moreover, the logarithmic scale in Figures~\ref{fig:bfs-time} and~\ref{fig:pr} compresses the apparent variance in runtime. Each platform in Fig.~\ref{fig:pr} has a relative standard deviation between $1/4$ and $1/2$ that of the same system executing SSSP.

The difficulty in comparing iteration counts for PageRank underscores an important challenge for any comparison of graph processing systems. The assumptions under which the various platforms operate can have a dramatic effect on the program. For example, the GAP Benchmark Suite can be recompiled to store weights as integers or floating-point values. This may affect performance in addition to runtime behavior in cases where weights like $0.2$ are cast to $0$. Similarly, how a graph is represented in the system (e.g., weighed or directed) may have performance and algorithmic implications but is not always readily apparent.

\subsection{Scalability}
\begin{figure}[htb]
	\centering
		% trim=left lower right upper
		\includegraphics[width=\linewidth, trim=0 18pt 18pt 12pt, clip]{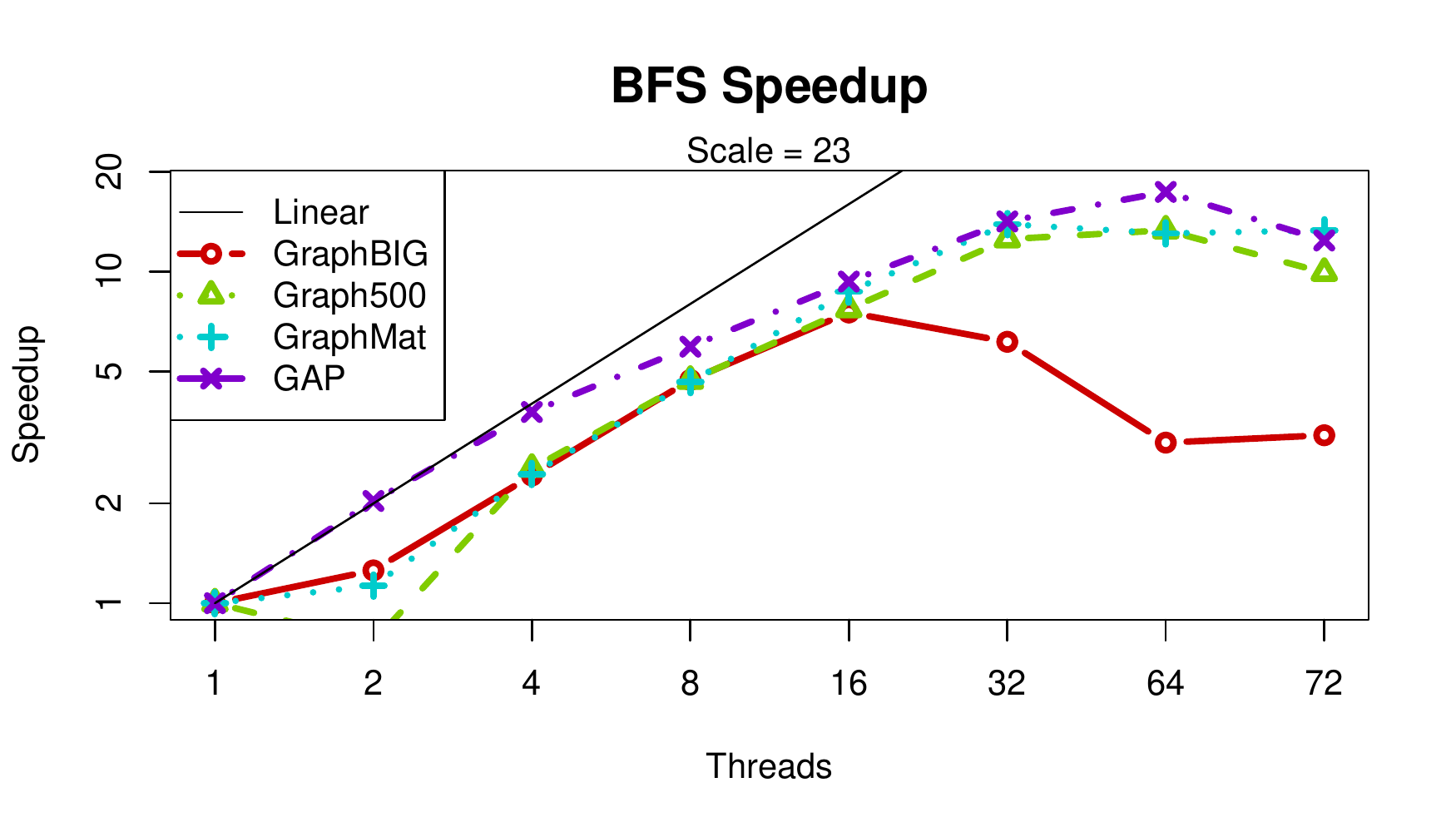}
		\caption{Speedup of BFS for a scale-23 graph, black solid line represents ideal speedup. Both axes are logarithmic with an exception at 72 threads for readability. }
	\label{fig:bfs-speedup}
\end{figure}

\begin{figure}[htb]
	\centering
		\includegraphics[width=\linewidth, trim=0 18pt 18pt 12pt, clip]{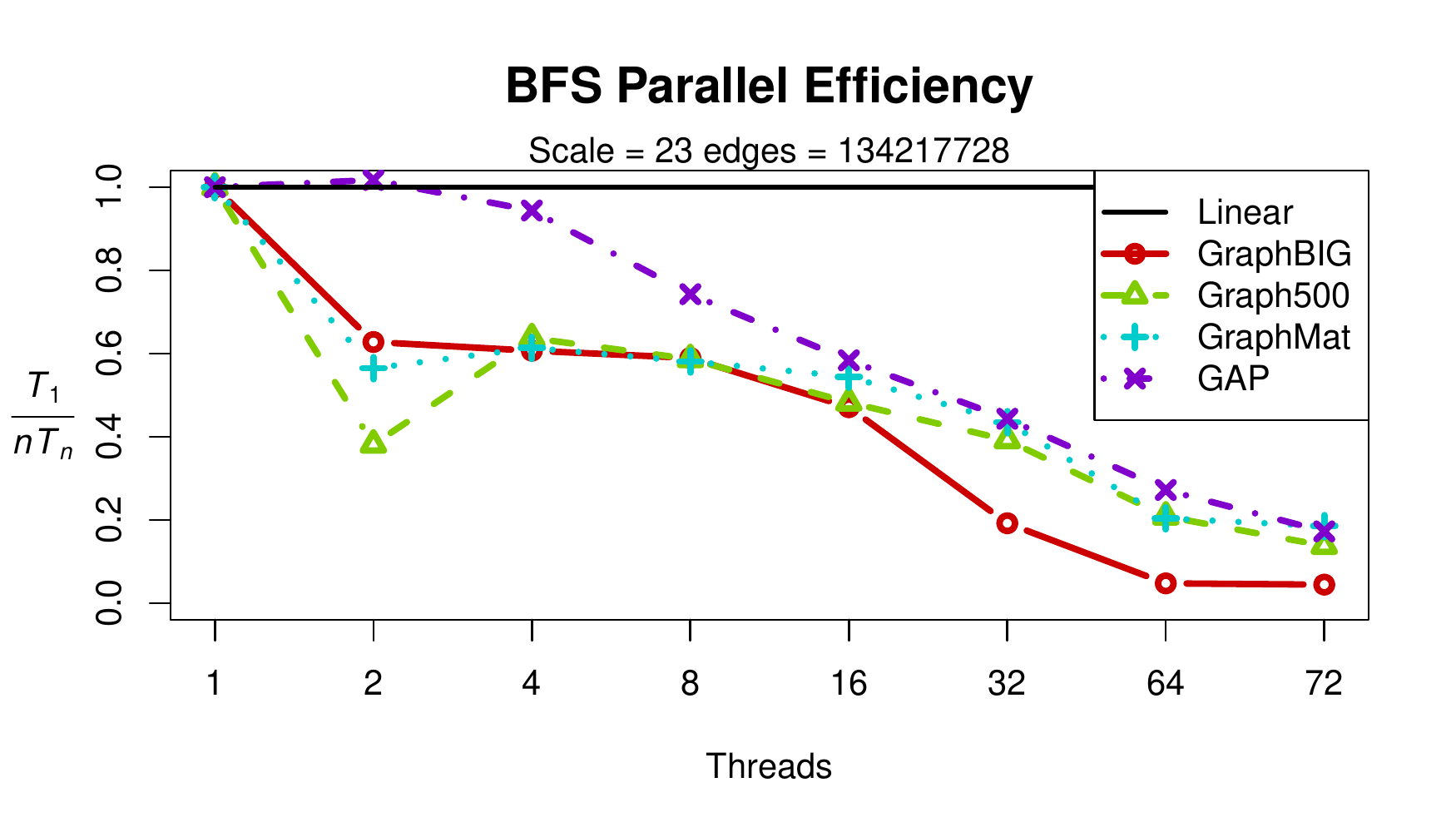}
	\caption{Parallel efficiency for a scale-23 graph. Graph500 dips below 1 because it is slower for $2$ threads than for 1. $T_1$ is the serial time, $n$ is the number of threads, and $T_n$ time with $n$ threads.}
	\label{fig:bfs-efficiency}
\end{figure}

Figures~\ref{fig:bfs-speedup} and~\ref{fig:bfs-efficiency} illustrate the parallel strong efficiency of the BFS algorithm in different packages on a Kronecker graph of scale $23$. The parallel speedup shown in Fig.~\ref{fig:bfs-speedup} is computed as $T_1 / T_n$ where $T_1$ is the sequential time and $T_n$ is the execution time on $n$ threads. Because of timing considerations, only four trials were run for these experiments.

Figure~\ref{fig:bfs-efficiency} shows the parallel efficiency,  $T_1 / (nT_n)$ for different implementations of BFS. Ideal efficiency is defined as $T_n = T_1/n$ and is the horizontal line near the top of Fig.~\ref{fig:bfs-efficiency}. 

These plots show generally poor scaling for this size problem, a challenge facing most parallel graph algorithms. In addition, $2^{23}$ vertices is a relatively small graph by today's standards and thus library designers focus on scalability for larger graphs. Another limitation may be the ability for OpenMP to efficiently handle such a large number of threads per machine. GCC version 4.8.5 supports OpenMP version 3.1 but the most recent release of OpenMP is 4.5.

Overall, GAP is the most scalable with GraphMat close behind for larger threads and even slightly beating GAP at 72 threads. While care was taken to ensure there was no other load on the system when the experiments were performed, the efficiency for two threads is surprisingly low for the Graph500. Because the Graph500 spends a shorter amount of time executing in general (there is no file I/O performs multiple BFS passes one right after another), it is more sensitive to spikes in CPU usage.

\subsection{Real-World Datasets}
\begin{figure}
	\includegraphics[width=\linewidth, trim=0 18pt 6pt 0pt, clip]{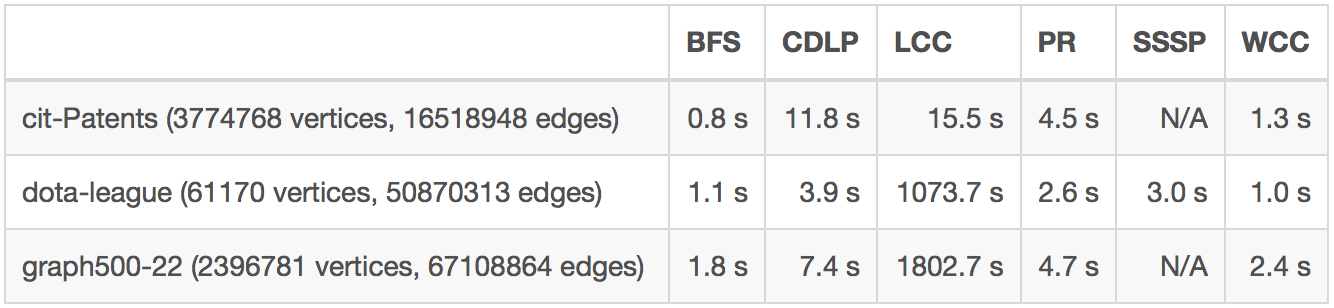}
	\caption{Real-world and synthetic experiments using Graphalytics on GraphBIG using 32 threads. Graphalytics outputs one HTML page per software package.}
	\label{fig:graphalytics-html}
\end{figure}

\begin{figure}
	\centering
	% trim=left lower right upper
	\includegraphics[width=\linewidth, trim=0 18pt 6pt 0pt, clip]{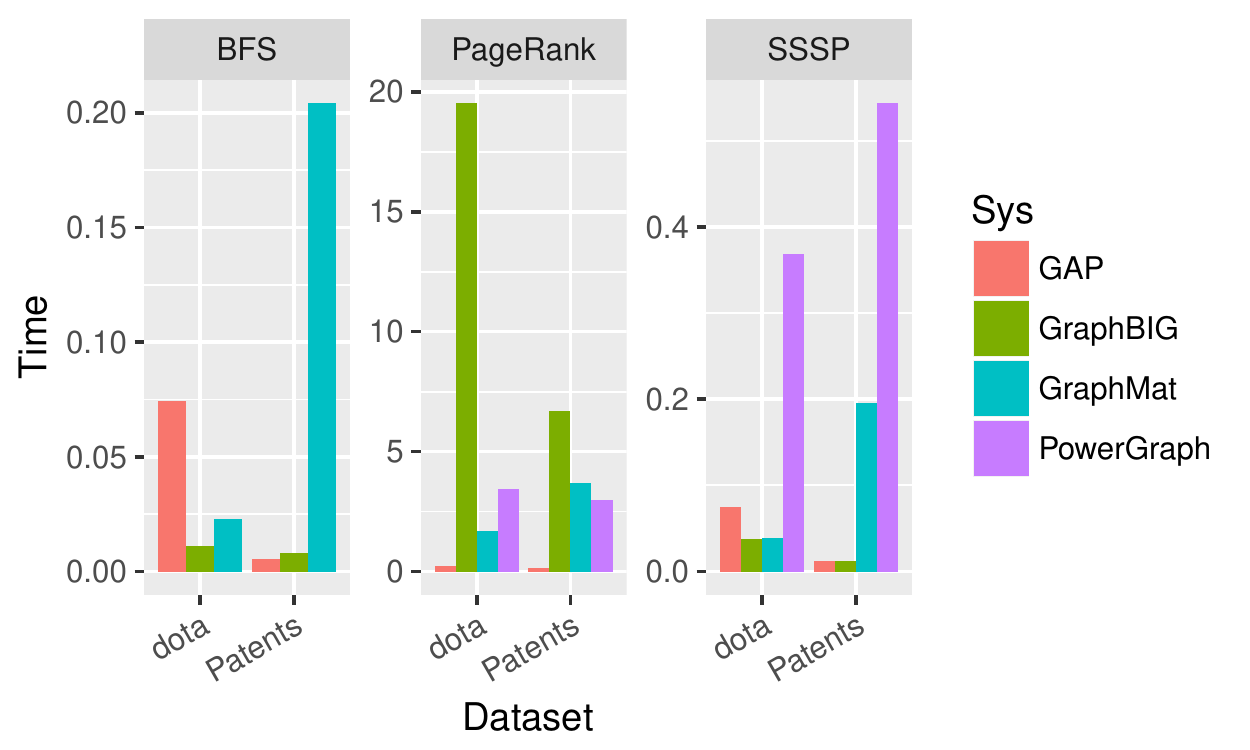}
	\caption{Real world experiments using \emph{easy-parallel-graph-\textasteriskcentered}. The leftmost plot does is missing Powergraph because Powergraph does not provide BFS.}
	\label{fig:epg-realworld}
\end{figure}

In addition to the data presented in Table~\ref{tab:graphalytics}, one may want to know the cost of performing additional computations on a graph once the data structures have been built. Figure~\ref{fig:epg-realworld} compiles average performance results for \emph{easy-parallel-graph-\textasteriskcentered}. We contrast this with Graphalytics which outputs an HTML page resulting from a single trial. We display a partial screenshot in \figurename~\ref{fig:graphalytics-html}.

Comparison of these two datasets in \figurename~\ref{fig:epg-realworld} brings up some surprising variations in the data. For example, PowerGraph is faster for SSSP. This could be because of the efficient edge-cut partitioning scheme in place on PowerGraph which can more efficiently deal with the high degree vertices present on the denser Dota-League graph. However, this comes with a significant overhead; PowerGraph is slower for SSSP than the other platforms. GraphBIG seems to have the widest variation; it is by far the slowest for PageRank but is the fastest for the Dota-League BFS, beating out even GAP which uses a more modern algorithm for BFS called Direction-optiming BFS \mbox{\cite{Beamer:2012:DOBFS}}. One reason for this lack of performance from GAP is our lack of tuning; we use the default parameterization of $\alpha = 15$ and $\beta = 18$, which may not be optimal for all graphs. One possible explanation for the poor performance of GraphMat compared is the overhead of the sparse matrix operations. These may pay off for larger datasets and we see good performance on the denser Dota-League dataset for GraphMat across all algorithms.

\subsection{Power and Energy Consumption}
We use the Performance Application Programming Interface (PAPI)~\mbox{\cite{Browne:2000:PAPI}} to gain access to Intel's Running Average Power Limit (RAPL), which provides a set of hardware counters for measuring energy usage. We use PAPI to obtain average energy in nanojoules for a given time interval. We modify the source code for each project to time only the actual BFS computation and give a summary of the results in Table~\ref{tab:power}. In our case, the fastest code is also the most energy efficient, although with this level of granularity we could detect circumstances where one could make a tradeoff between energy and runtime.

\begin{table*}
	\caption{The data are generated using a Kronecker graph with a scale of 22. The program is executed with 32 threads. Sleeping Energy refers to the power (in Watts) consumed during the \texttt{unistd} C \texttt{sleep} function, multiplied by the Time row. Essentially, this measures the energy that would have been consumed when the CPU and memory are (nearly) idle. The increase over sleep is the ratio of the first and third columns. These are all averaged over 32 roots.}
	\centering
	\begin{tabular}{l|r|r|r|r}
		\large{\emph{easy-parallel-graph\mbox{-\textasteriskcentered}}}	&	GAP  &    Graph500 & GraphBIG & GraphMat \\ \hline
		Time (s) &  0.01636 & 0.01884 & 1.600 & 1.424 \\
		Average Power per Root (W) & 72.38 & 97.17 & 78.01 & 70.12 \\
		Energy per Root (J) &	1.184 & 1.830 & 112.213 & 111.104 \\
		Sleeping Energy (J) & 0.4046  & 0.4660 & 39.591 &  35.234 \\
		Increase over Sleep & 2.926 & 3.928 & 2.834 & 3.153
	\end{tabular}
	\label{tab:power}
\end{table*}

\begin{figure}
	\centering
	\begin{minipage}{0.78\linewidth}
		\includegraphics[width=\linewidth, trim=0 36pt 18pt 0, clip]{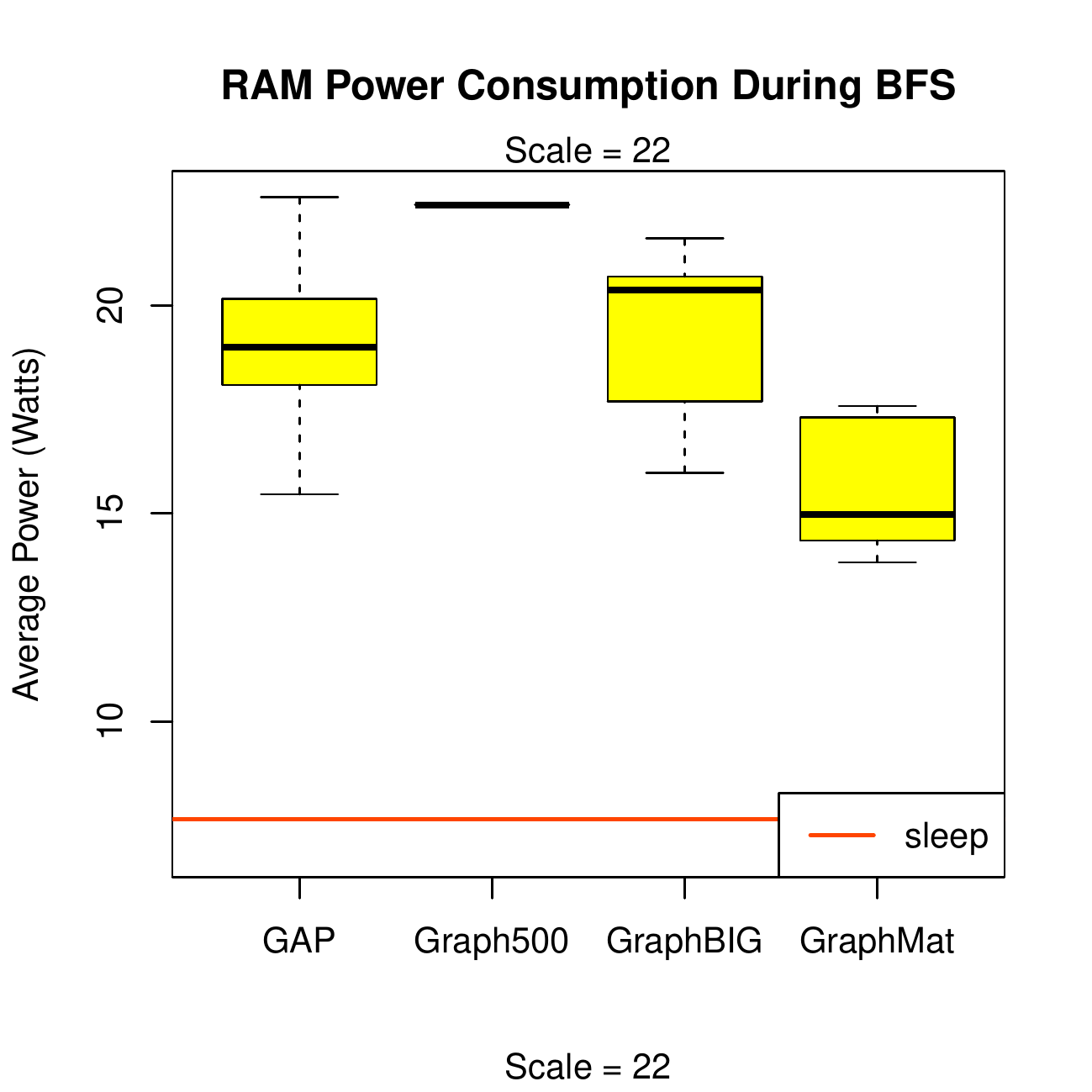}
	\end{minipage}
	\begin{minipage}{0.78\linewidth}
		\includegraphics[width=\linewidth, trim=0 36pt 18pt 0, clip]{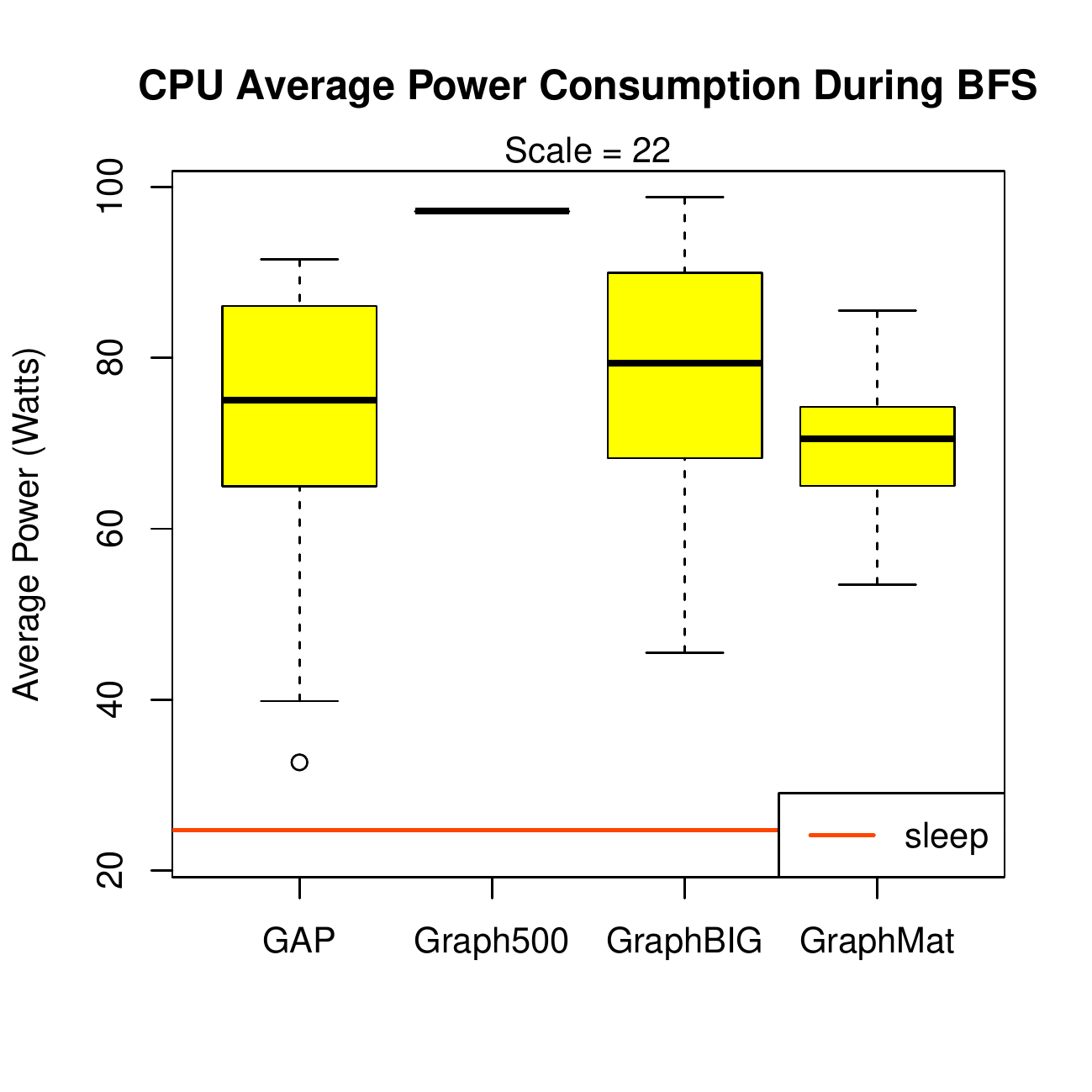}
	\end{minipage}
	\caption{Similar to Fig.~\ref{fig:bfs-time} we plot RAM and CPU Power Consumption for each of the 32 roots. Since the Graph500 runs multiple roots per execution, we only get a single data point. The baseline was computed by monitoring power consumption while a program containing only one call to the C \texttt{unistd} library function \texttt{sleep(10)} (ten seconds).}
	\label{fig:power}
\end{figure}

RAPL also allows the measurement of DRAM power, the results of which are shown in Fig.~\ref{fig:power}. The right plot describes the second row of Table~\ref{tab:power} in more detail. We notice a smaller spread of RAM power consumption, but still a noticeable difference. GraphMat exhibits the lowest average RAM power consumption (but not the shortest execution time). This information can be useful when choosing an algorithm to use in limited power scenarios, where a slower algorithms that will not exceed the power cap is preferred to a faster one that may exceed it.

\begin{figure}
	\begin{lstlisting}[language=C, frame=single]
#ifdef POWER_PROFILING
power_rapl_t ps;
power_rapl_init(&ps);
power_rapl_start(&ps);
#endif
	<region of code to profile>
#ifdef POWER_PROFILING
power_rapl_end(&ps);
power_rapl_print(&ps);
#endif
	\end{lstlisting}
	\begin{lstlisting}[language=make,  basicstyle=\footnotesize]
CFLAGS += -I$(PAPI)/include -DPOWER_PROFILING=1
LDLIBS += -L$(PAPI)/lib
LDLIBS += -Wl,-rpath,$(PAPI)/lib -lpapi
	\end{lstlisting}
	\caption{The changes required to monitor energy usage for a C or C++ file and its corresponding Makefile, given the papi header file is included and \$PAPI is the directory where PAPI is installed.}
	\label{lst:RAPL}
\end{figure}

Listing~\ref{lst:RAPL} shows the simplicity of adding RAPL profiling to existing code. Once the source code is modified, it is compiled with our provided header file and linked with a small library. Alternatively, there may be scenarios where this approach is not feasible. For example, in some cases, if the user does not have root privileges s/he may not be able to access the model specific registers storing the energy usage. One alternative is to use TAU paired with PAPI to instrument the source code, and if the counters are available on the host CPU then the power and energy can still be measured.

\section{Future Work}\label{sec:fw}
%We found that GraphMat is only decently performant, but scales to more threads. One explanation is that GraphMat's underlying computation model (sparse matrix operations), paired with the increased overhead of GraphMat's doubly-compressed sparse row (DSCR) graph representation, is better suited to larger-scale graphs.
% Furthermore, sparse matrix operations have been well studied so achieving good scalability is more realistic [cite]

Our choice of algorithms is based on the common subset of methods analyzed in prior work and implemented in the selected software packages. The standardization of graph algorithm building blocks (graph kernels) is being developed by the Graph BLAS Forum~\mbox{\cite{Mattson:2013:graphblas}}). Once this standardization is finalized there is motivation from both library designers and performance analyzers to implement and profile each kernel. Regardless, algorithms like triangle counting and betweenness centrality are widely implemented but not supported by either Graphalytics nor \emph{easy-parallel-graph\mbox{-\textasteriskcentered}}. 

Advances in parallel SSSP and BFS contain parameterizations ($\Delta$ for SSSP and $\alpha$ and $\beta$ for BFS) which affects performance depending on graph structure. These are provided in GAP. We plan to add some level of heuristic parameter tuning as performed in \mbox{\cite{Beamer:2012:DOBFS}} to the next iteration of our framework to take advantage of these algorithmic advances.

Graphalytics encountered circumstances with the more computationally expensive algorithms fail~\mbox{\cite{Iosup:2016:Graphalyticstech}}, so determining whether an algorithm will finish given a particular machine, input size, runtime limit, and resources is an important unanswered question we plan to pursue further.

Even though most software packages represent graphs using CSR format, the implementation details differ across packages. There may be significant performance differences among the various packages between using directed or undirected, or weighted and unweighted graphs.
% Our framework abstracts away much of the data formatting problem, but issues such as whether a framework supports a given dataset is a complex task. For example, GraphMat currently cannot convert weighted graphs to unweighted and does not support zero-indexed graphs.

One overarching issue with these software packages is the speed at which they change. On one end of the spectrum, the code base of PowerGraph was made closed-source in 2015. Simple fixes such as increasing the maximum number of threads per process and updating dependency URLs are not merged and as a result there is much duplicated effort among more than 400 forks of PowerGraph. On the other end, GraphMat's update to version 2.0 is in general incompatible with version 1.0 and as such has not yet been adopted by Graphalytics or our framework. One potential solution would be to add each framework to a package management system such as Spack \mbox{\cite{Gamblin:2015:Spack}} which supports packaging of high performance computing software.

With respect to power and energy profiling, while our current implementation supports measurements based on PAPI's interface to RAPL, which is only available on Intel platforms, the interface is simple and easy to adapt to other platforms in future without requiring PAPI support. In particular, fine-grained measurements provided through potentially available custom hardware~\mbox{\cite{naecon15}} can be enabled through the same interface.

\section{Conclusion}
We have presented a tool called \emph{easy-parallel-graph\mbox{-\textasteriskcentered}}, where \textasteriskcentered $\ = $ installer, comparator, performance analyzer, dataset homogenizer, or performance visualizer. This framework can analyze both power consumption and the two fundamental phases of execution: data structure construction and algorithm execution. While the comparison presented here can help one choose among alternatives for the selected packages and algorithms, the problem of selecting a graph processing framework for a given large-scale problem remains far from simple. Furthermore, increasing hardware heterogeneity demands performance analysis be easily repeatable on the target architecture. Because of this, our framework is designed to be easy to use and we make our code freely available at \url{https://github.com/HPCL/easy-parallel-graph} to encourage further experimentation.

Overall, the GAP Benchmark Suite is the best-performing system across all chosen datasets and in general the most scalable for the graphs used in this study (at most $2^{23}$ vertices). In addition, GAP is the most recent project and also the most robust in terms of the input graphs accepted.  Powergraph and Graphmat are designed for larger graphs (it is possible to run in distributed mode for both) so the overhead of these frameworks may dominate for smaller problem sizes. Two more potentially important considerations are cost and portability: GraphMat requires the Intel compiler collection which may be cost-prohibitive for some users. Despite the GAP Benchmark Suite being the overall best performing benchmark, it was not the best performing in every case so we unsurprisingly recommend the more comprehensive comparison using all five benchmarks offered by \emph{easy-parallel-graph-\textasteriskcentered}.

\bibliographystyle{IEEEtranS}
\bibliography{drp}
\end{document}